\documentclass[intlimits,twoside,a4paper]{article}

\usepackage{amsmath,amssymb}
\usepackage{graphicx}
\usepackage{wrapfig}

\usepackage[T2A]{fontenc}
\usepackage[cp1251]{inputenc}

\usepackage[eqsecnum]{cmpj3}

\issue{2017}{20}{4}{43601}
\doinumber{10.5488/CMP.20.43601}

\title[Self-assembly of linear triblock copolymer]%
{The effect of  asymmetry of the coil block on self-assembly
in ABC coil-rod-coil triblock copolymers}
\author[X.-G. Han, H.-H. Meng, Y.-H. Ma, S.-L. Ouyang]{X.-G. Han\refaddr{label1}\footnote{xghan0@163.com}\,, H.-H.
Meng\refaddr{label1}, Y.-H. Ma\refaddr{label1}, S.-L. Ouyang\refaddr{label2}}
\addresses{
\addr{label1} School of Science, Inner Mongolia University of
Science and Technology, Baotou 014010, China 
\addr{label2} Inner Mongolia Key Laboratory for Utilization of Bayan Obo Multi-Metallic
Resources: Elected State Key Laboratory, Inner Mongolia University
of Science and Technology, Baotou 014010, China }

\date{Received January 25, 2017, in final form June 5, 2017}
\begin{document}

\maketitle

\begin{abstract}
Using the self-consistent field approach, the effect of asymmetry of
the coil block on the microphase separation is focused in ABC
coil-rod-coil triblock copolymers. For different
 fractions of the rod block $f_{\text B}$, some stable structures are observed,
i.e., lamellae, cylinders, gyroid, and core-shell hexagonal lattice,
and the phase diagrams are constructed. The calculated results show
that the effect of the coil block fraction $f_{\text A}$ is dependent on
$f_{\text B}$. When $f_{\text B}=0.2$, the effect of asymmetry of the coil block is
similar to that of  the  ABC flexible triblock copolymers; When
$f_{\text B}=0.4$, the self-assembly of ABC coil-rod-coil triblock copolymers
behaves like rod-coil diblock copolymers under some condition. When
$f_{\text B}$ continues to increase, the effect of asymmetry of the coil
block reduces. For $f_{\text B}=0.4$, under the symmetrical and rather
asymmetrical conditions, an increase in the interaction parameter
between different components leads to  different transitions
between cylinders and lamellae. The results indicate some remarkable
effect of the chain architecture on self-assembly, and can
provide the guidance for the design and synthesis of copolymer
materials.
\keywords self-consistent field theory, rod-coil block copolymer,
self-assembly
\pacs 61.25.Hp, 64.75.+g, 82.60.Fa
\end{abstract}

\section{Introduction}
Block copolymers are macromolecules composed of chemically distinct
subchains or blocks. The distinct blocks tend to phase separate,
whereas the covalent bonds between distinct blocks prevent a
macroscopic separation. The competition of these two opposing trends
leads to the formation of various ordered phases \cite{Haml2004}.
From a fundamental point of view, the ordering in block copolymers
provides an ideal paradigm for the study of self-organization in
soft condensed matter \cite{Haml1998}. From a technological point of
view, the rich and fascinating ordered structures from block
copolymers have found applications ranging from thermoplastic
elastomers and high-impact plastics to pressure-sensitive adhesives,
and so on. Compared with coil-coil block copolymers, rod-coil block
copolymers have received much attention due to their abilities to
self-assemble smaller scale structures in the field of novel
functional and biomimetic materials \cite{Leco2001,Losi2004}.
Especially, the rod-coil block copolymers containing conjugated rod
building blocks offer opportunities for engineering electronic and
optical devices \cite{Ball2006,Wete2006,Dai2007}.

In contrast to coil-coil block copolymers, rod-coil diblock copolymers
have rich phase behaviors. Rod-coil
block copolymers, as the simplest rod-coil block copolymers,
self-assembled into ordered structures, e.g., zigzag
lamellae \cite{Chen1995}, stripes \cite{Radz1997},
honeycombs \cite{Lee2001}, and hollow spherical and cylindrical
micelles \cite{Jene1998}. Compared with AB
diblock copolymers, ABC triblock copolymers have more independent
parameters controlling their phase behavior. The phase behavior of
the two-component rod-coil block copolymers is controlled mainly by
three parameters: the volume fraction of the rod block $f_{\text A}$, the
Flory-Huggins interaction between different blocks $\chi_{\text{AB}}$, and
the total degree of polymerization of the copolymer $N$. For ABC
triblock copolymers, the number of parameters increases to six,
including three interaction parameters $\chi_{\text{AB}} $, $\chi_{\text{BC}} $,
and $\chi_{\text{AC}} $; two independent volume fractions $f_{\text A}$ and $f_{\text B}$
and the total degree of polymerization of the copolymer $N$.
This increased number of molecular variables will impose varieties
and complexities on the self-assembly of the rod-coil block
copolymers, meanwhile, leading to a great model system for
engineering of a large number of intriguing nanostructures.

Some studies have centered on the microphase separation of
rodlike-coil three-component
 block copolymers. Experimentally,
 Lee and coworkers \cite{Lee1998,Oh2006} studied ABC coil-rod-coil triblock
molecules, focusing the fraction of coil block on the order-order
transition \cite{Lee1998}. Zhou et al. \cite{Zhan2005} reported an
ABC coil-coil-rod triblock containing liquid crystal polymer as the
rod block, and found the formation of a liquid crystalline phase
when the rod block achieved a certain length. Chang et al.
\cite{Chan2011} found that the rod-rod interaction lead to an
order-disorder transition, and the lamellae appeared for
$f_{\text{rod}}\simeq0.2$ in ABC rod-coil-coil triblock copolymers.
Theoretically, Xia et al. studied the self-assembly of
 ABC rod-coil-coil and coil-rod-coil triblock copolymers \cite{Xia2010},
 constructing the phase $f_{\text{rod}}$ vs. $\chi_ N$ phase diagram.
 Recently, Li et al. concentrated on
the order-to-order phase transitions in coil-worm-coil
triblock copolymers, suggesting that the tuning of the flexibility
parameter provides a promising approach to design
the resulting microphase-separated structures in semiflexible
copolymer melts \cite{Li2016}. Although some literatures
reported the effect of the fraction of end block on
self-assembly, a systematic theoretical study on the molecularly
asymmetry of the copolymers containing rigid blocks   does not
exist so far.

Due to the flexibility of adjusting system parameters, theory
and simulation provide an ideal approach to explore the phase
behavior of block copolymers. Mayes and Olvera de la Cruz \cite{Gao2013}, as
well as Dobrynin and Erukhimovich \cite{Li2014} investigated flexible ABA triblock
copolymers in the weak-segregation limit, providing evidence that
molecular asymmetry can have a profound effect on both
order-disorder and order-order transitions. Later on, Matsen 
used the self-consistent field theory \cite{Mats1994} (SCFT) to show that
such molecular asymmetry can alter microdomain dimensions and
order-order transitions. Recently, Woloszczuk et al. employed
on-lattice Monte Carlo simulations to examine the phase behavior of
molecularly flexible asymmetric  ABA copolymers in the limit of superstrong
segregation, wherein interstitial micelles composed of the minority A
endblock were observed to arrange into two-dimensional hexagonal
arrays along the midplane of B-rich lamellae. These studies have
demonstrated that molecular asymmetry is an important factor for
self-assembly of block copolymers. In this work, we will
systematically study the effect of the asymmetry of the coil
block on self-assembly in the  rod-coil block
copolymers.

SCFT  can be used to study rod-coil block copolymer systems.
Pryamitsyn and Ganesan
 introduced Maier-Saupe
theory \cite{Prya2004,Jian2013,Gao2013,Li2014} to really account for
the aligning interactions between rods, and the  anisotropy
interaction  between rods presented considerable computational
challenges. Some other theoretical models did not consider
anisotropic interactions between the rods. For instance, Chen et~al. \cite{Chen2006}, as well as Li and Gersappe \cite{Li2001} performed
lattice-based SCFT simulations provided with computational
advantage. The work of Chen et al. was the first time
that SCFT predicted the hexagonal cylinder phase for rod-coil block
copolymers. Subsequently Chen et al.  used the SCFT lattice
model to study the self-assembly of
two-component rod-coil block copolymers \cite{Chen2007,Chen2008,Xia2008,Xia2009} and  studied firstly in
theory three-component ones \cite{Xia2010}.  In
this work, using SCFT lattice model, we concentrate on the effect
of the volume fraction of the coil block on structure transition in the
 ABC coil-rod-coil triblock copolymers, providing for different rod block fractions.
The paper is organized as follows: In section~\ref{sec2}, we present the
lattice model SCFT of coil-rod-coil triblock copolymers. The
approach employs a two-stage relaxation procedure to evolve a
system as rapidly as possible to a free-energy minimum similar to
the literature \cite{Chen2006}. In section~\ref{sec3}, the calculated
results are presented.
These results are compared with those of previous works, especially
with flexible ABC triblock copolymers. A brief conclusion will be
given in section~\ref{sec4}.

\section{Theory}\label{sec2}
We consider $n$ linear ABC coil-rod-coil triblock copolymers in a
lattice, where the degree of polymerization of each chain is $N$. The
A, B and C segments  have the same size and each segment occupies
one lattice site. Thus, the total number of the lattice sites $N_{\text L}$
equals $n(f_{\text A}+f_{\text B}+f_{\text C})N$. The system partition function can be
expressed as
\vspace{-3mm}
\begin{equation}
Z=\sum_{{r(j,s)},{\alpha(j,s)}}\left(\frac{1}{N_{\text L}^n}\frac{1}{z^n}\prod_{j=1}^{n}\prod_{s=2}^{N}\lambda
_{r_{j,s}-r'_{j,s-1}}^{\alpha _{j,s}-\alpha'
_{j,s-1}}\right)\exp{\left(-\frac{U}{k_{\text B}T}\right)}.
\end{equation}

Here, $r_{j,s}$ and $\alpha_{j,s}$ denote the position and bond
orientation of the $s$-th segment of the $j$-th copolymer,
respectively. $r'$ denotes the nearest neighboring site of $r$.
$\alpha_{j,s}$ can choose all of the possible bond orientations,
which depends on the selected lattice model. $z$ denotes lattice
coordination number. The transfer matrix lambda depends only on the
chain model used. This paper adopts an inflexion chain model. For a
coil subchain,
\begin{eqnarray}
\lambda _{r_{j,s}-r'_{j,s-1}}^{\alpha _{j,s}-\alpha'
_{j,s-1}}=\left\{
\begin{array}{ll}
0, & \alpha_{j,s}=-\alpha'_{j,s-1}\,,\\
1/(z-1), &\text{otherwise}.\\
\end{array} \right.
\end{eqnarray}
For a rigid subchain,
\begin{eqnarray}
\lambda _{r_{j,s}-r'_{j,s-1}}^{\alpha _{j,s}-\alpha' _{j,s-1}}=
\left\{
\begin{array}{ll}
1, & \alpha_{j,s}=\alpha' _{j,s-1}\,,\\
0, & \text{otherwise}.\\ 
\end{array} \right.
\end{eqnarray}
Following the scheme of Scheutjens and Leemakers \cite{Leer1988}, the
end-segment distribution function $G^{\alpha _{s}}(r,s|1)$ presents
a statistical weight of all possible configurations staring from
segment 1, which can be located in any position within the lattice,
ending at segment $s$ at site $r$, and satisfies the following
recurrence relation:
\begin{equation} G^{\alpha _{s}}(r,s|1)=G(r,s)\sum_{r_{s-1}^{\prime
}}\sum_{\alpha' _{s-1}}\lambda _{r_{s}-r_{s-1}^{\prime }}^{\alpha
_{s}-\alpha' _{s-1}}G^{\alpha _{s-1}}(r^{\prime },s-1|1).
\label{free}
\end{equation}
For all the values of $\alpha _{1}$, the initial condition is
$G^{\alpha _{1}}(r,1|1)=G(r,1)$, $G(r,s)$ is the weight factor of
the free segment. It is expressed as $G(r,s)=\exp[-\omega
_{\beta}(r_{s})]$,  $s\in \beta$ $(\beta=\text{A},\text{B},\text{C})$. Another end-segment
distribution function $G^{\alpha _{s}}(r,s|N)$ satisfies the
following recurrence relation:
\begin{equation}
G^{\alpha
_{s}}(r,s|N)=G(r,s)\sum_{r_{s+1}^{\prime }}\sum_{\alpha'
_{s+1}}\lambda _{r_{s+1}^{\prime }-r_{s}}^{\alpha _{s+1}-\alpha'
_{s}}G^{\alpha _{s+1}}(r^{\prime },s+1|N).
\end{equation}
With the initial condition $G^{\alpha _{_N}}(r,N|N)=G(r,N)$ for all
the values of $\alpha_{N}$.

The free energy functional of $F$ (in the unit of $k_{\text B}T$ ) in
canonical ensemble is defined by
\begin{align}
F&=\frac{1}{z}\sum_{rr'}[\chi_{\text{AB}}\phi_{\text A}(r)\phi_{\text B}(r')+\chi_{\text{AC}}\phi_{\text A}(r)\phi_{\text C}(r')+\chi_{\text{BC}}\phi_{\text B}(r)\phi_{\text C}(r')]-\sum_r[\omega_{\text A}(r)\phi_{\text A}(r)\nonumber\\
&\quad+\omega_{\text B}(r)\phi_{\text B}(r)+\omega_{\text C}(r)\phi_{\text C}(r)]-\sum_r{\xi(r)[1-\phi_{\text A}(r)-\phi_{\text B}(r)-\phi_{\text C}(r)]}-n\ln{Q}\,,
\end{align}
where
\begin{equation}
Q=\frac{1}{N_{\text L}}\frac{1}{z}\sum_{r_N}\sum_{\alpha_N}G^{\alpha^{N}}(r,N|1).
\end{equation}
Here, $\chi_{\text{AB}}$, $\chi_{\text{AC}}$, $\chi_{\text{BC}}$ are the Flory-Huggins
interaction parameters between different species. The $\phi_k(r)$ is
the volume fraction field of block specie $k$, which is independent
of the individual polymer configuration, and $\omega_k(r)$ is the
chemical potential field conjugated to $\phi_k(r)$. The $\xi(r)$ is
the potential field that ensures the incompressibility of the
system, also known as a Lagrange multiplier.

Minimizing the free energy functional $F$ with respect to
$\phi_{\text A}$, $\phi_{\text B}$, $\phi_{\text C}$, $\omega_{\text A}$, $\omega_{\text B}$,
$\omega_{\text C}$ and $\xi(r)$ lead to the following SCFT equations:
\begin{equation}
\omega_{\text A}(r)=\frac{1}{z}\sum_{r'}\chi_{\text{AB}}\phi_{\text B}(r')+\frac{1}{z}\sum_{r'}\chi_{\text{AC}}\phi_{\text C}(r')+\xi(r),
\end{equation}
\begin{equation}
\omega_{\text B}(r)=\frac{1}{z}\sum_{r'}\chi_{\text{BC}}\phi_{\text C}(r')+\frac{1}{z}\sum_{r'}\chi_{\text{AB}}\phi_{\text A}(r')+\xi(r),
\end{equation}
\begin{equation}
\omega_{\text C}(r)=\frac{1}{z}\sum_{r'}\chi_{\text{AC}}\phi_{\text A}(r')+\frac{1}{z}\sum_{r'}\chi_{\text{BC}}\phi_{\text B}(r')+\xi(r),
\end{equation}
\begin{equation}
\phi_{\text A}{(r)}+\phi_{\text B}{(r)}+\phi_{\text C}{(r)}=1,
\end{equation}
\begin{equation}
\phi_{\text A}(r)=\frac{1}{N_{\text L}}\frac{1}{z}\frac{n}{Q}\sum_{s\in{\text{A}}}\sum_{\alpha_s}\frac{G^{\alpha_s}(r,s|1)G^{\alpha_s}(r,s|N)}{G(r,s)}\,,
\end{equation}
\begin{equation}
\phi_{\text B}(r)=\frac{1}{N_{\text L}}\frac{1}{z}\frac{n}{Q}\sum_{s\in{\text{B}}}\sum_{\alpha_s}\frac{G^{\alpha_s}(r,s|1)G^{\alpha_s}(r,s|N)}{G(r,s)}\,,
\end{equation}
\begin{equation}
\phi_{\text C}(r)=\frac{1}{N_{\text L}}\frac{1}{z}\frac{n}{Q}\sum_{s\in{\text{C}}}\sum_{\alpha_s}\frac{G^{\alpha_s}(r,s|1)G^{\alpha_s}(r,s|N)}{G(r,s)}\,.
\end{equation}

In our calculations, the real-space method is implemented to solve
the SCFT equations in a cubic lattice with periodic boundary
conditions \cite{Chen2006,Xia2010}. The calculations begin from the
initial potential fields generated by random functions, and stop
when the free energy of the system changes within a tolerance of
$10^{-8}$. By comparing the system free energies from different
initial potential fields, we obtained  stable phases in the system
of ABC coil-rod-coil triblock copolymers.

\section{Result and discussion\label{sec3}}

\begin{figure}[!t]
\centering
\includegraphics[width=0.3\textwidth]{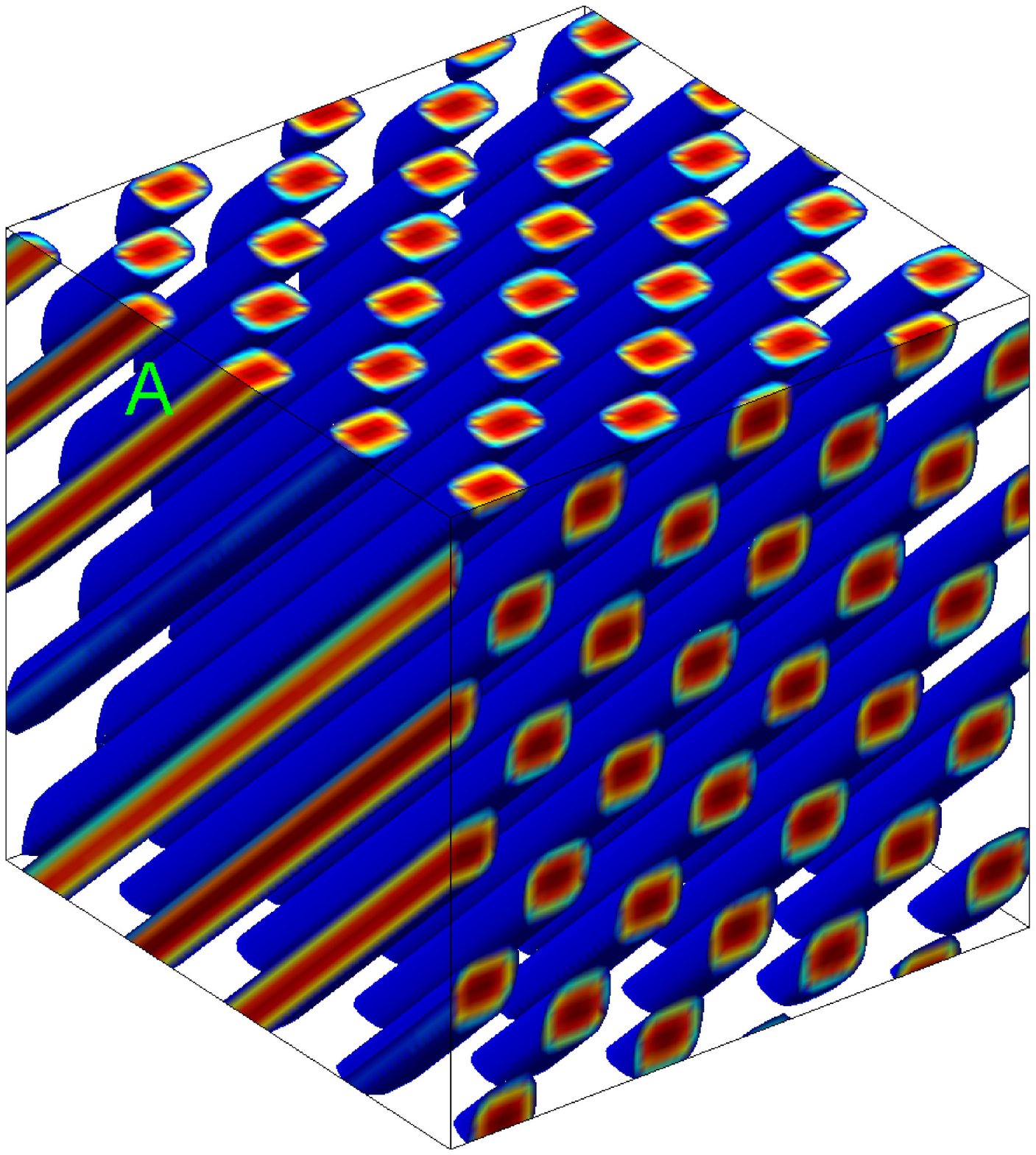}\hspace{3mm}\includegraphics[width=0.3\textwidth]{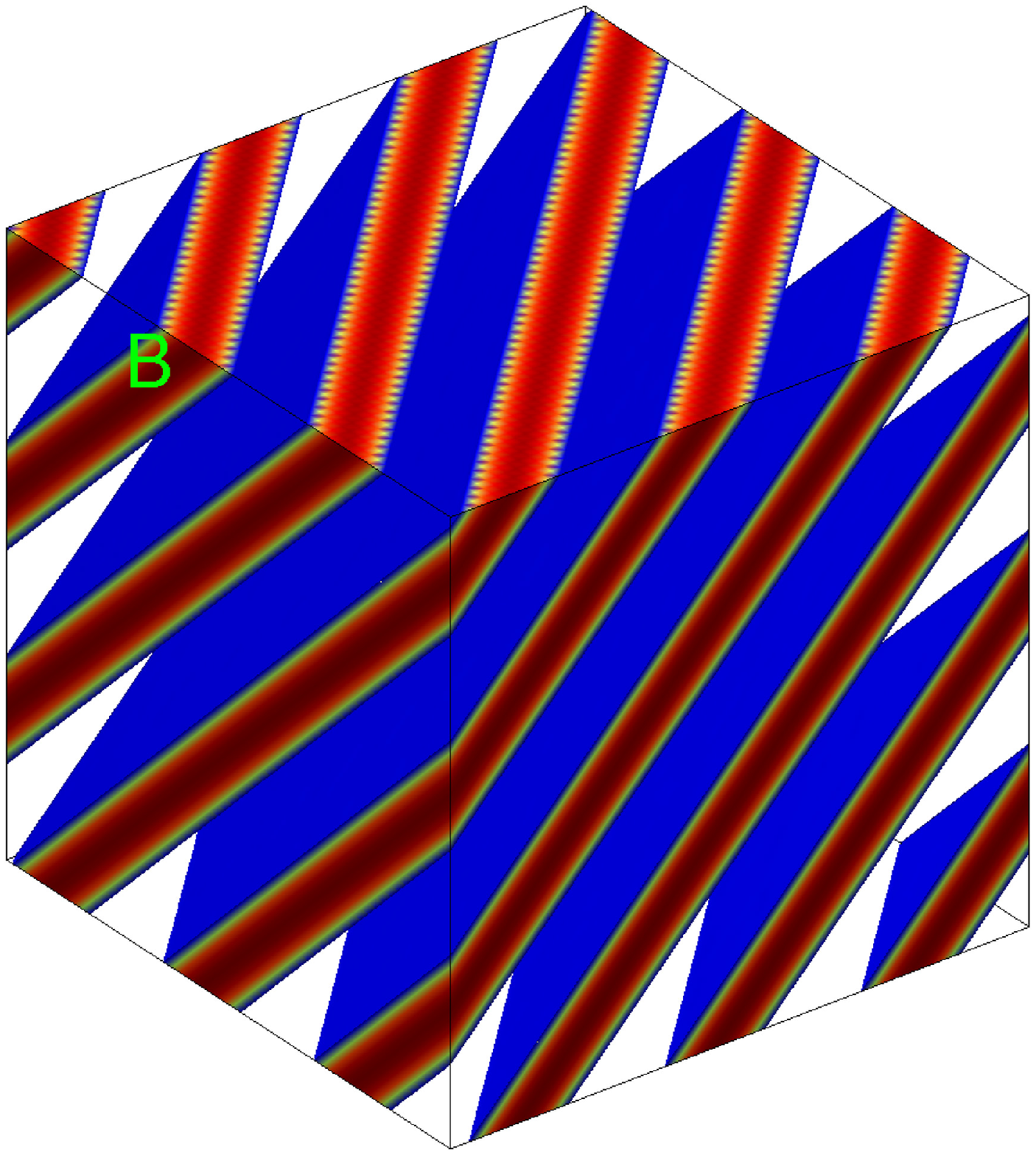}\hspace{3mm}
\includegraphics[width=0.3\textwidth]{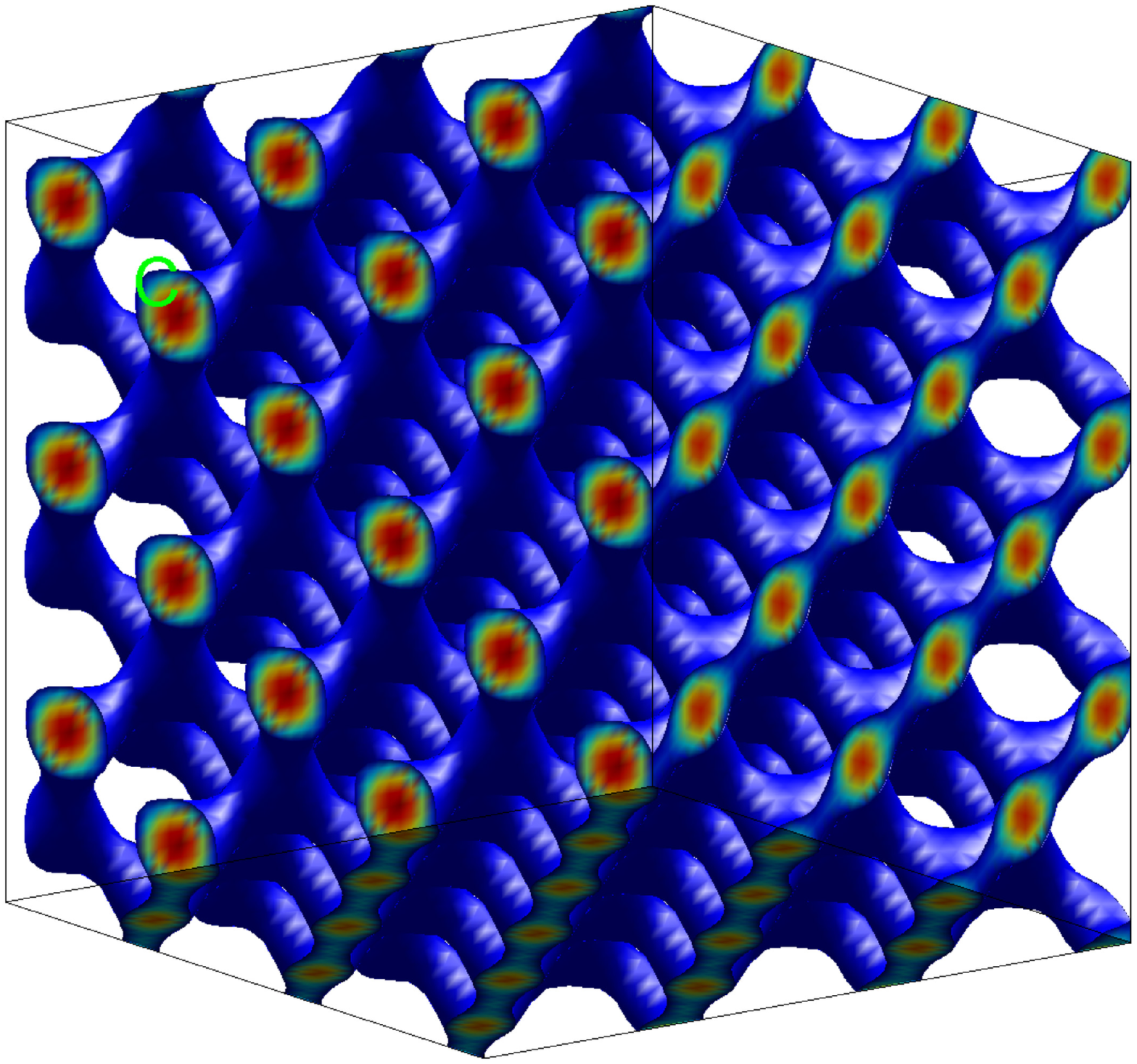}\\
\includegraphics[width=0.3\textwidth]{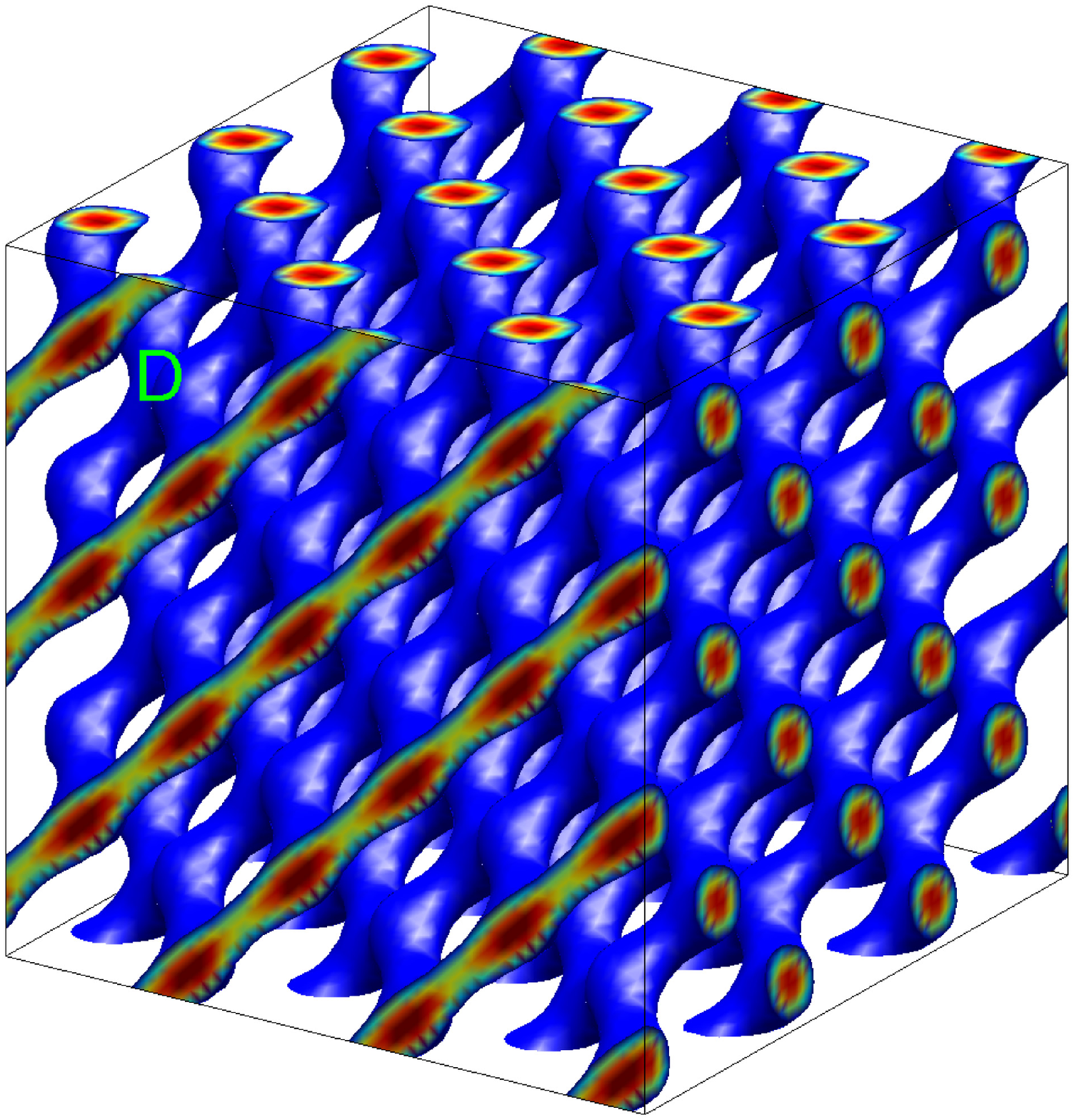}\hspace{3mm}
\includegraphics[width=0.3\textwidth]{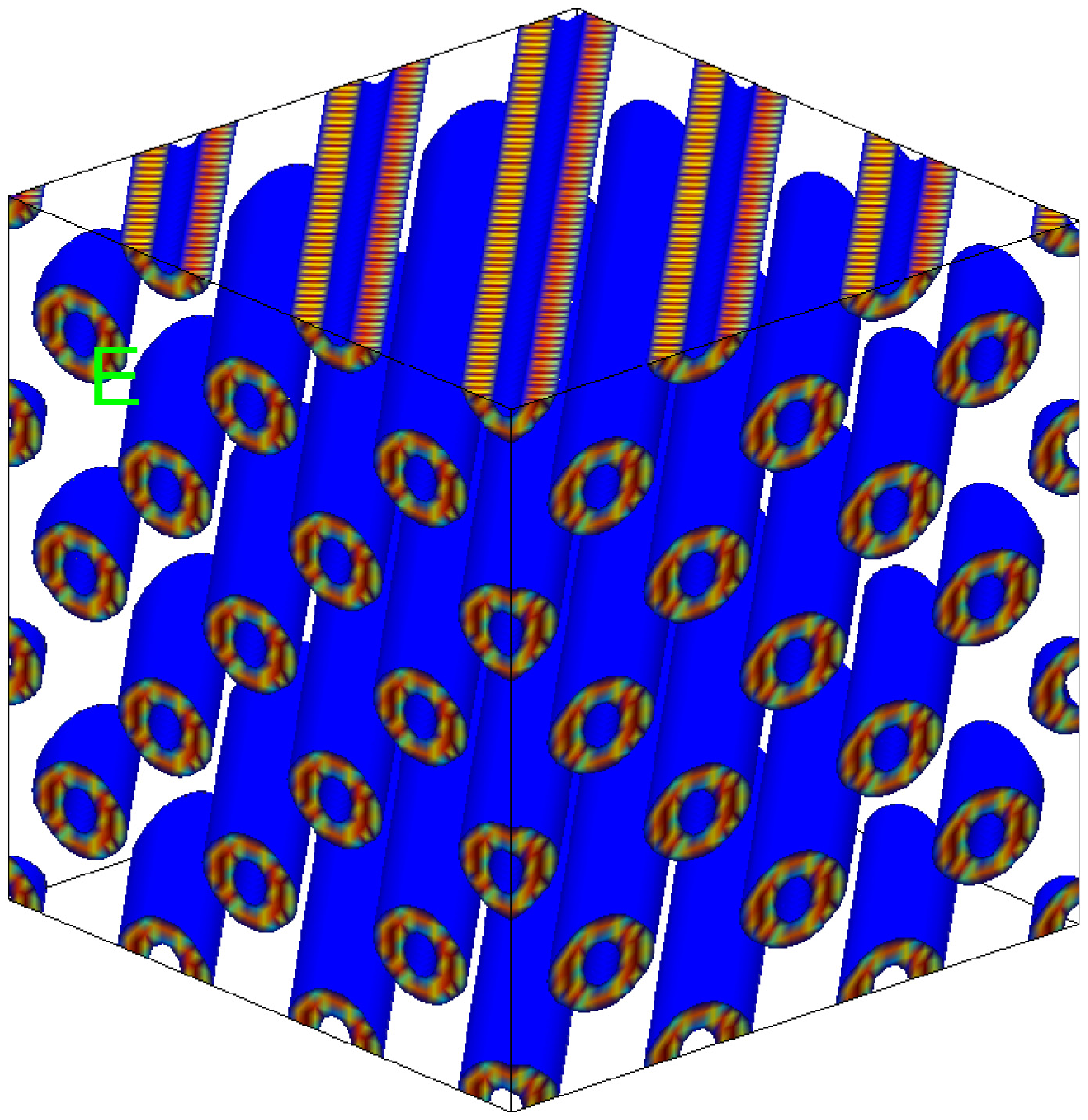}
\caption{(Color online) The self-assembled structures of  ABC
coil-rod-coil triblock copolymers: (a) cylinder (the components A and C are eliminated for clarity);
(b) lamella (the components A and C are eliminated for clarity);
(c) gyriod (the components A and B are eliminated for clarity);
(d) perforated lamella phase (the components A and C are eliminated for clarity);
(e) core-shell hexagonal lattice (the components A and C are eliminated for clarity). \label{phadia1}}
\end{figure}

\begin{figure}[!t]
\centering
\includegraphics[width=0.45\textwidth]{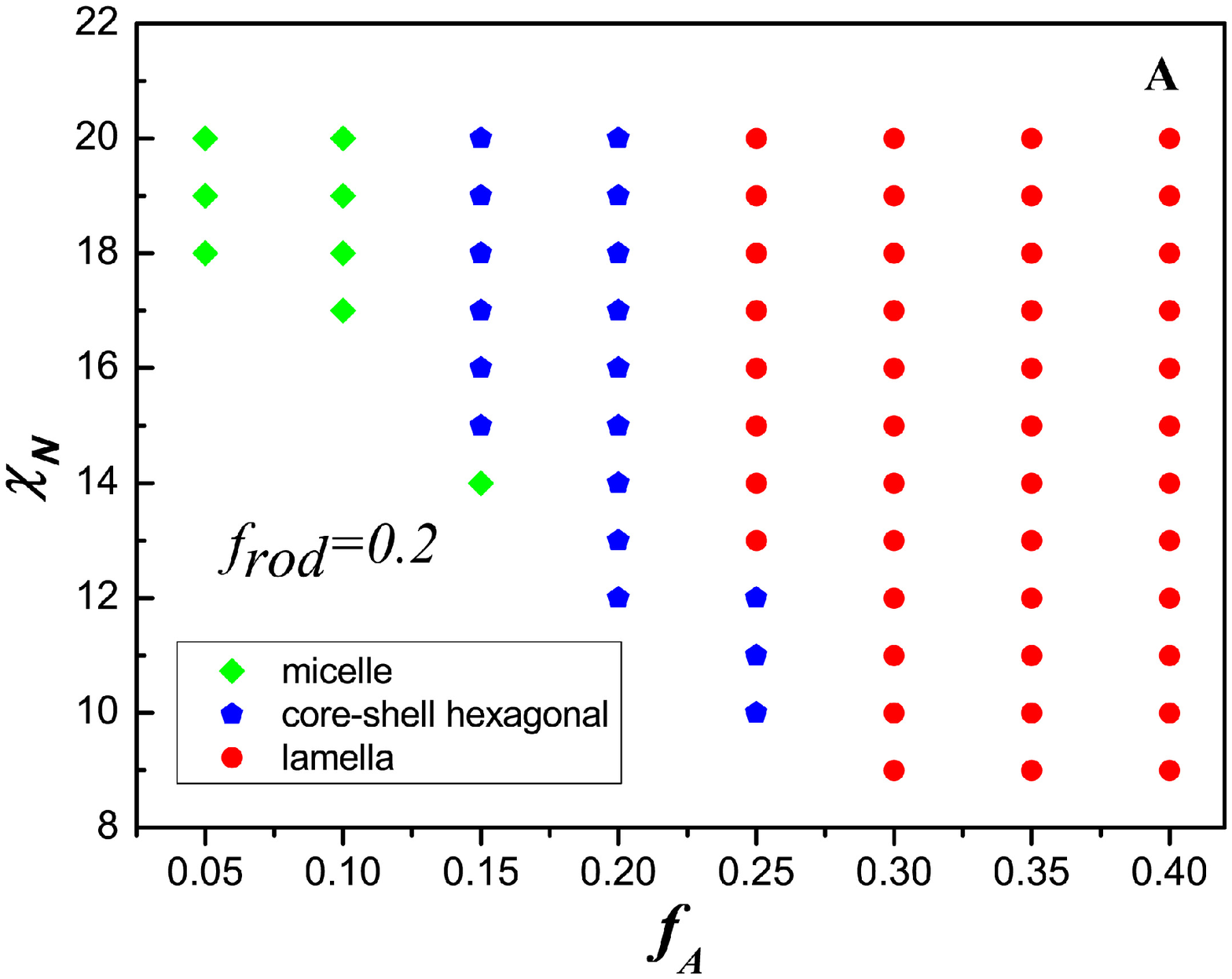}
\includegraphics[width=0.45\textwidth]{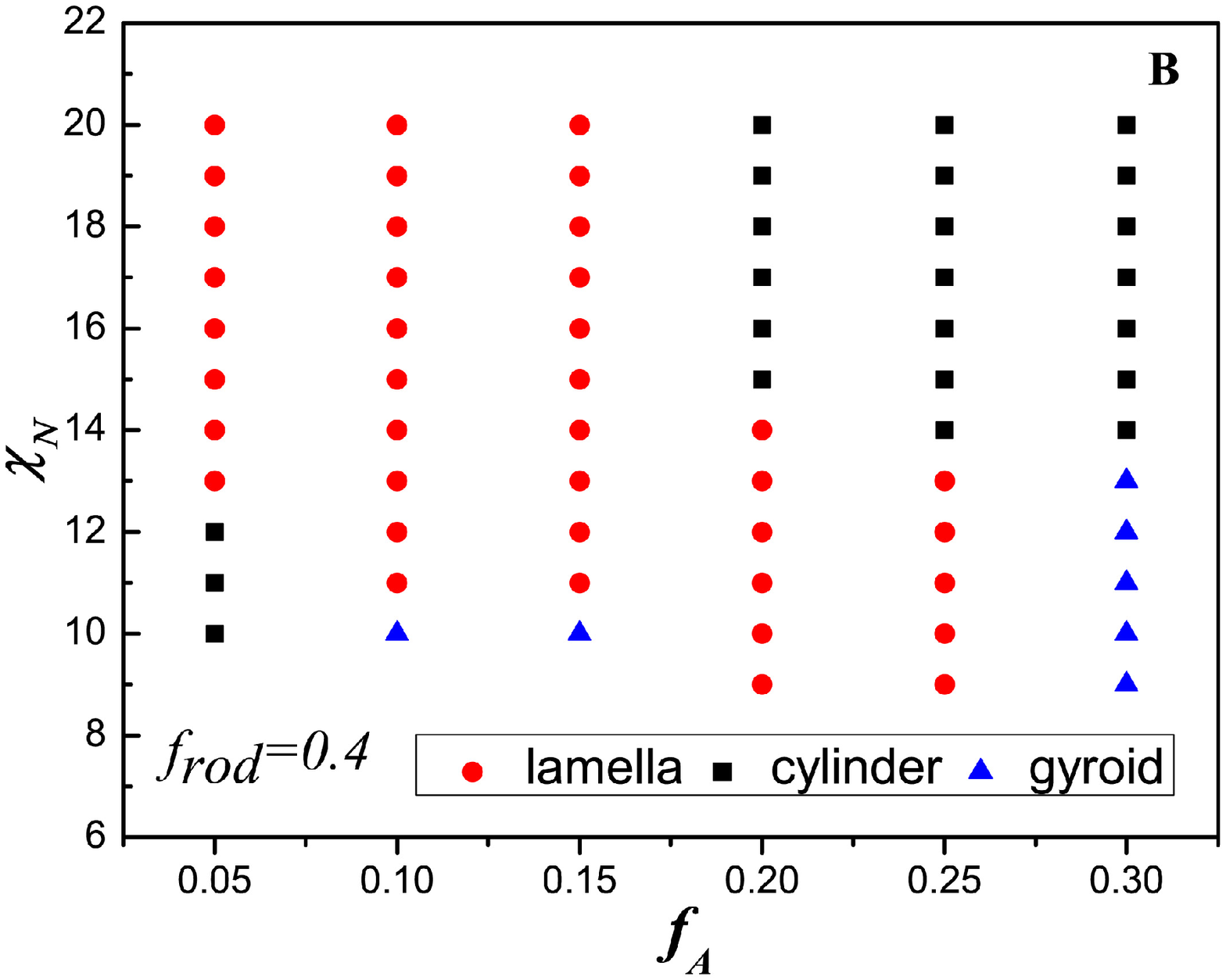}
\includegraphics[width=0.45\textwidth]{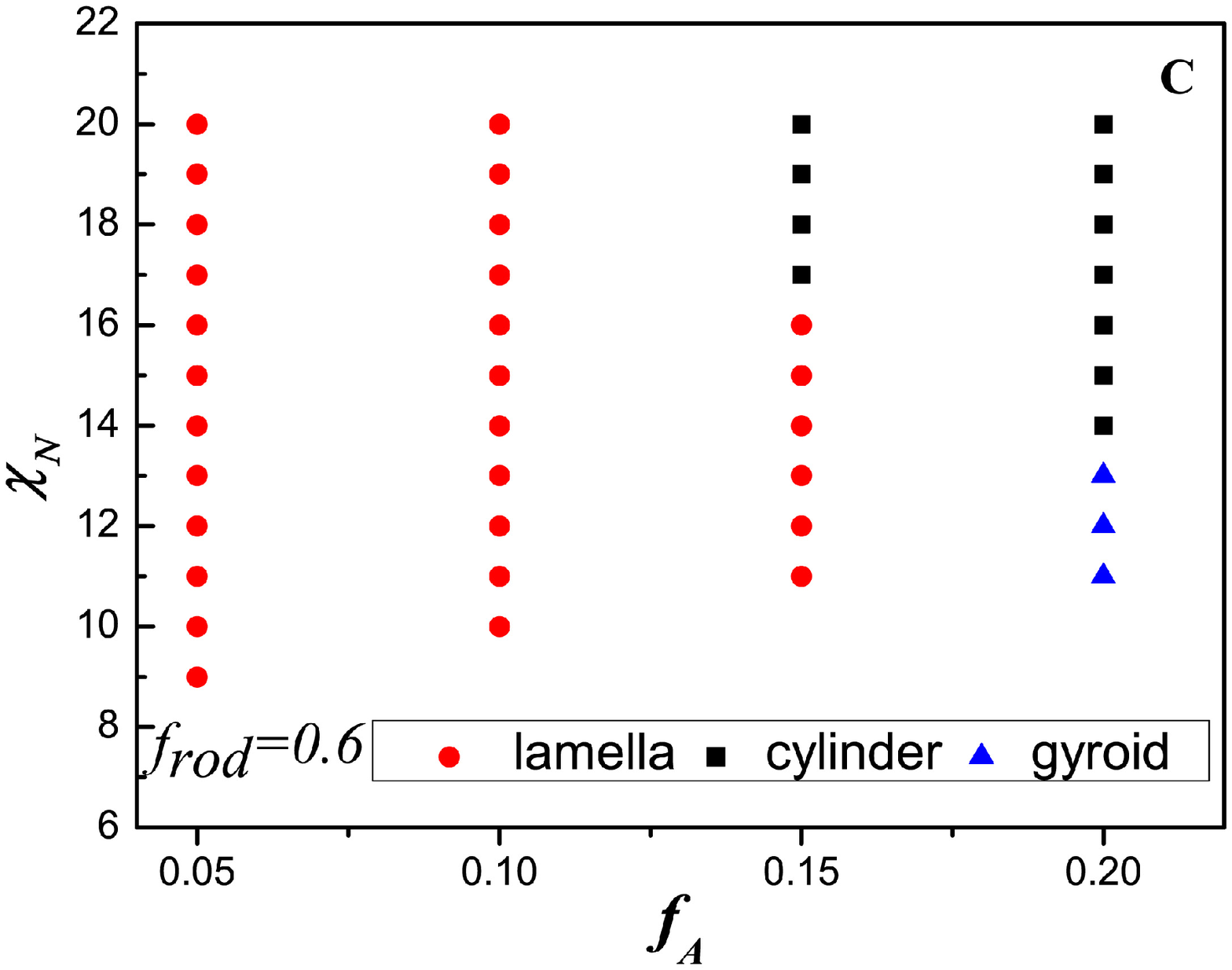}
\caption{(Color online) The phase diagram for the ABC coil-rod-coil  triblock
copolymers with different length of the end A coil block when $
f_{\text B}=0.2, 0.4, 0.6$, respectively. \label{phadia2}}
\end{figure}

In our studies, we explore the effect of the asymmetry of the coil
block on structural transition in ABC
coil-rod-coil triblock copolymers. The interaction parameters
between  different species are the same, i.e., $\chi_{\text{AB}}=\chi
_{\text{BC}}=\chi_{\text{AC}}=\chi$, and the degree of the polymerization is
$N=20$. Our calculations are preformed in the lattices of
$N_{\text L}=60^{3}$ and $N_{\text L}=80^{3}$, and the emergence of the
self-assembled structures are not constrained by system size.  As
showed in figure~\ref{phadia1}, some stable structures are observed, i.e.,
cylinders, core-shell hexagonal lattice (CSH) phase, lamellae and
gyriod, and a perforated lamellae phase appears as a metastable
structure. The phase diagrams for  copolymers constructed by
$f_{\text A}$ versus $\chi N$ are shown in figure~\ref{phadia2} for $f_{\text B}=0.2, 0.4,
0.6$, and most of the regions of all the phase diagrams are occupied
by lamellar phase. This indicates that the tolerance of the lamellae to chain
architecture asymmetry  is the highest of these
stable phases in ABC triblock copolymers.

In the case of $f_{\text B}=0.2$, and when $f_{\text A}$ $(=0.05, 0.1)$ is small, among
 the three phase, i.e., micelles, CSH phase and lamellae, micelles are the most
 stable, and the A and B components coexist in the domain of a micelle.
 With an increase in $f_{\text A}$, CSH phase ($0.15\leqslant f_{\text A}\leqslant 0.20$), where
 the core and shell severally consist of A and B
 components, firstly becomes more stable, and then
lamellae ($0.25\leqslant f_{\text A}\leqslant 0.40$) are stable phase. For rod-coil diblock copolymers, at
$f_{\text B}=0.2$, lamellae are not observed, while micelles are observed. Therefore, the
existence of a third block leads to the
appearance of lamellae, which is confirmed by the flexible
 triblock copolymer \cite{Tang2004}. The CSH phase and lamellae also appear in
turn under the similar conditions in flexible ABC
ones, but the A and B component-mixed phase is not observed.
However, it is expected that micelles will change into CSH phase when
the interaction parameter is rather big. Therefore, for $f_{\text B}=0.2$, the
effect of the length of the end block on microphase separation in
ABC coil-rod-coil triblock copolymers is similar to that of the
flexible ones.

\begin{figure}[!t]
\centering
\vspace{-1mm}
\includegraphics[width=0.3\textwidth]{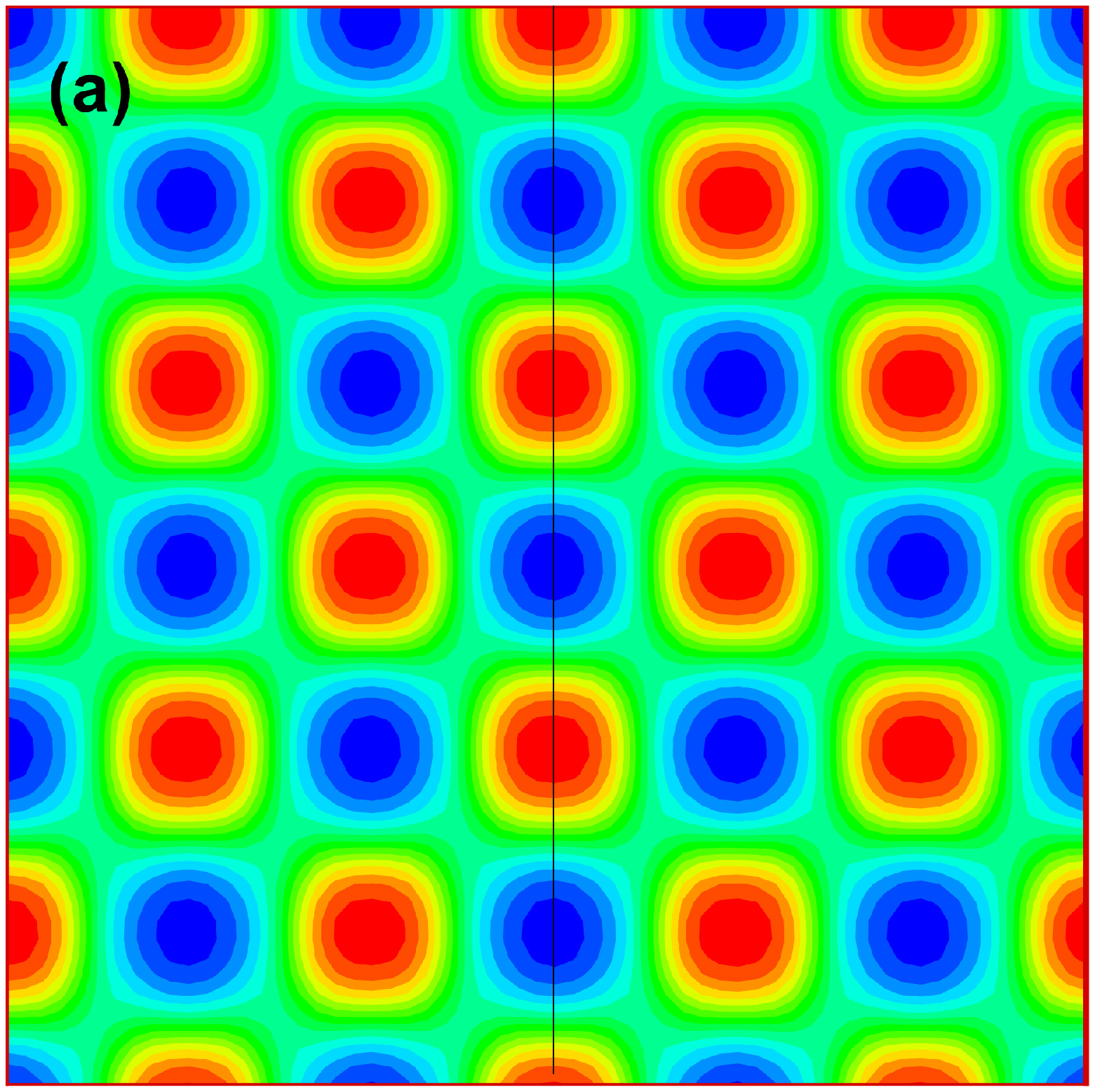}\qquad\includegraphics[width=0.3\textwidth]{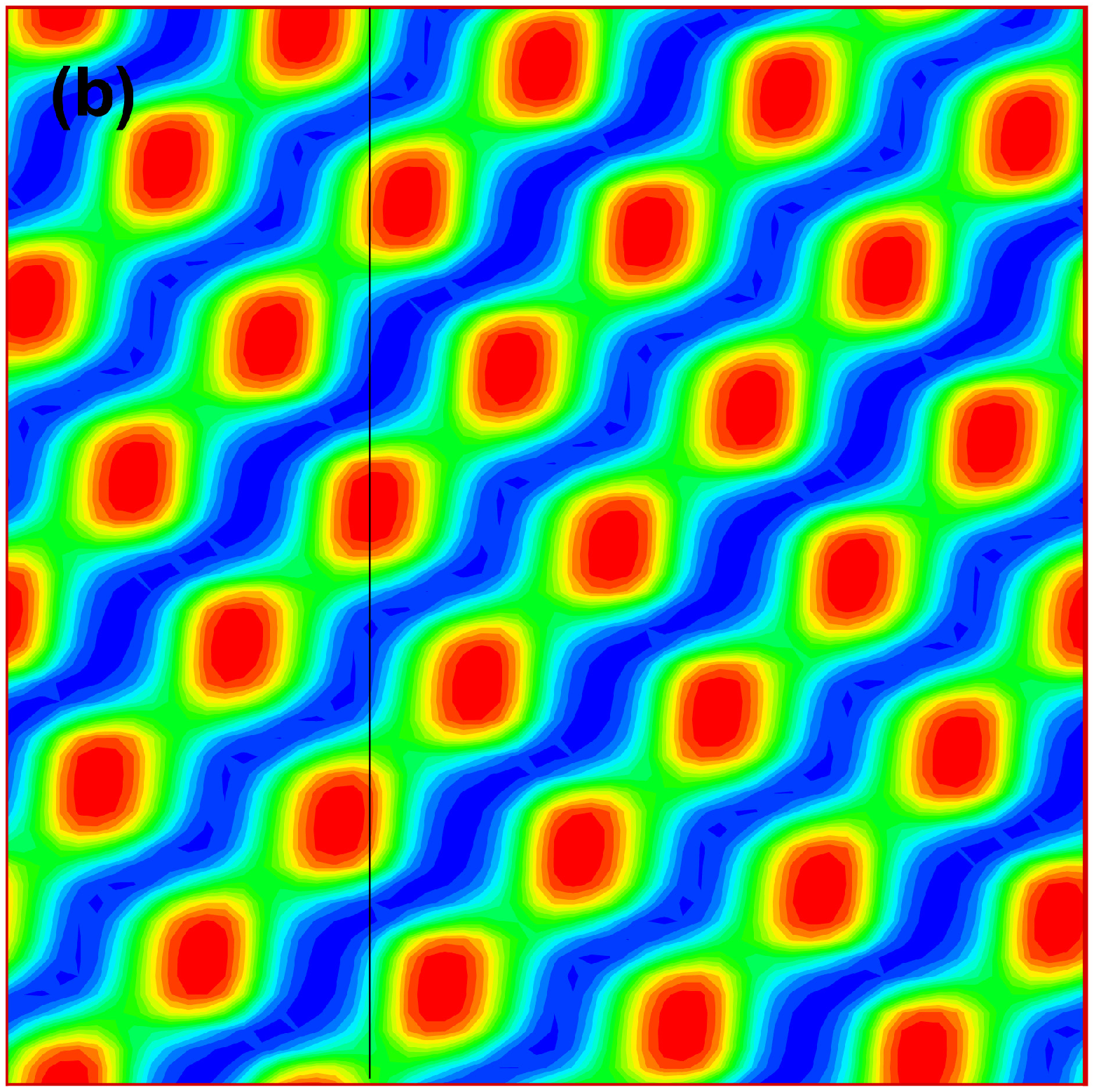}\vspace{3mm}\\
\includegraphics[width=0.55\textwidth]{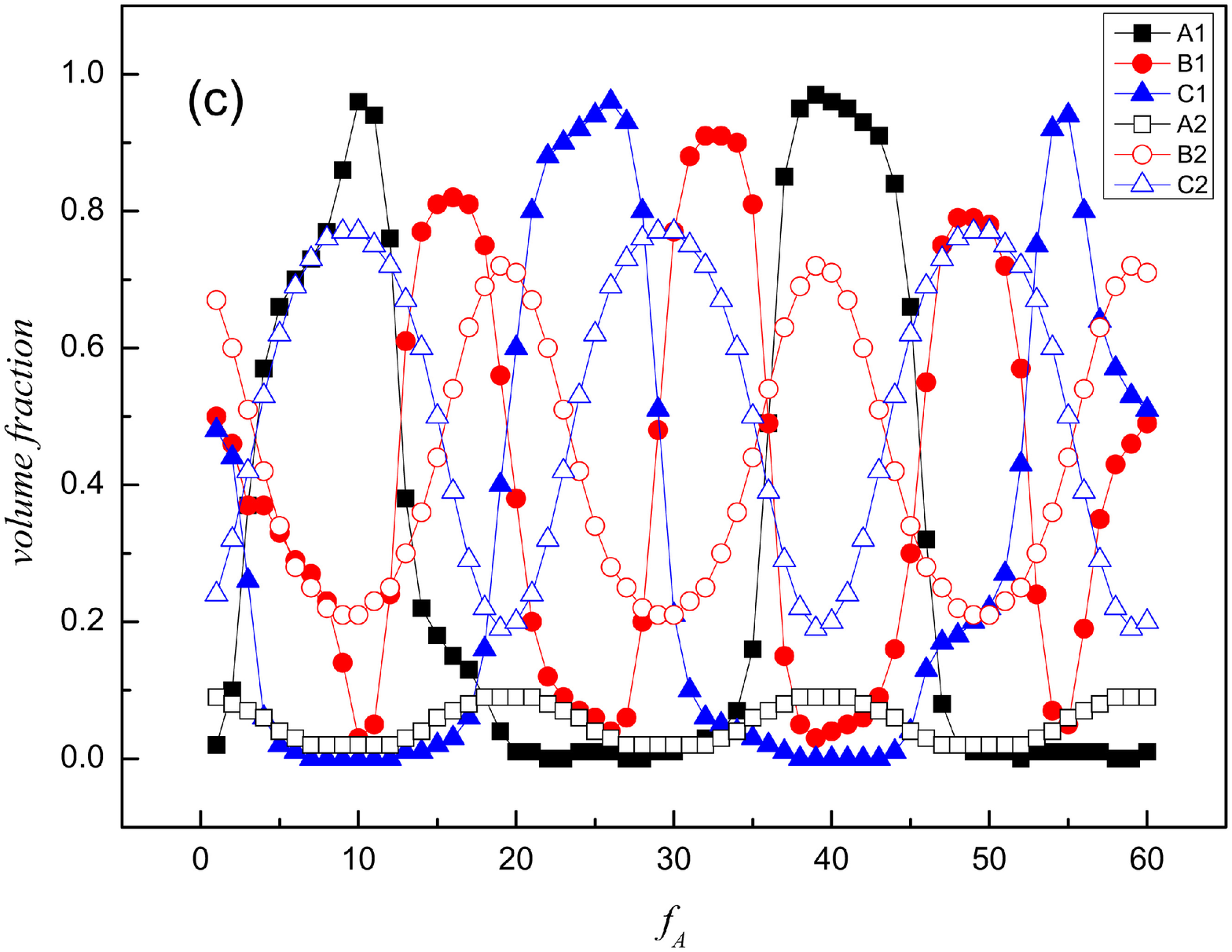}
\caption{\label{fig3} (Color online) Figures~(a) and (b) are  cross sections of the two
cylindrical structures with $f_{\text A}=0.05$ and $f_{\text A}=0.3$,
respectively. Figure~(c) indicates the density changes of three
components along the black line in the figures~(a) and (b), the
hollow and solid symbols correspond  to $f_{\text A}=0.05$ and
$f_{\text A}=0.3$, respectively.}
\end{figure}

\looseness=-1 For  $f_{\text B}=0.4$, when the two coil blocks are symmetrical, gyriod and cylinders can be
found, gyriod is more stable in the small parameter region of
$9\leqslant\chi N\leqslant13$. With an increase in $\chi N$, cylinders are
more stable than gyriod. With the appearance of the asymmetry of the
copolymers ($0.20\leqslant f_{\text A}\leqslant 0.25$), gyriod is replaced by lamellar structure. When $\chi
N\geqslant14$, the free energy of the lamellae is higher than that of
cylinders, and thus cylinders remain stable. When $f_{\text A}=0.15, 0.10$, the
cylinder structure is also replaced by lamellae into the larger  $\chi
N$ region, and near the order-disorder transition, the copolymers
self-assemble into gyriod again. When $f_{\text A}=0.05$, cylinders are observed  in a
small $\chi N$ region, which is different from the case of the large
$f_{\text A}$. When $\chi N\geqslant 12$, lamellae are still observed. In the
flexible ABC triblock copolymers, however, the three-color lamellae
do not appear when $f_{\text A}<0.1$. It is obvious that the appearance of
lamellae at $f_{\text A}=0.05$ results from the rigidity of the middle block
copolymer of coil-rod-coil triblock copolymer. Accordingly, the
chain stiffness is propitious to the occurrence of lamellae, which
is scrupulously discussed in bottlebrush block
copolymers \cite{Chre2016}. Furthermore, when $13\leqslant\chi N\leqslant16$,
perforated lamellar phase is observed as a metastable structure.
When $f_{\text A}$ becomes small, the gyriod and cylinders sequentially
appear for small $\chi N$, indicating that the self-assembly for the
asymmetric coil-rod-coil triblock copolymers tends to be that of the
coil-rod diblock copolymers, which is in detail discussed below.
It is noted that with an increasing $\chi
N$, the lamellar phase translates into cylinders for large $f_{\text A}$, and
vice versa at $f_{\text A}=0.05$. The effect of the length of the end A block is
concerned with the fraction of the middle rod block. For
$f_{\text B}=0.6$, the distribution of the ordered structures in a phase diagram
[see figure~\ref{phadia2}~(c)] is similar to the one of $f_{\text B}=0.4$ for large
$f_{\text A}$. It is shown that the effect of asymmetry of the end coil
block on self-assembly will reduce when $f_{\text B}$ continues to increase.

In order to clarify the difference of the above two transitions
between lamellae and cylinders for $f_{\text B}=0.4$, as shown in figure~\ref{fig3}, the
cylindrical phases for different $f_{\text A}$ are compared below. When
$f_{\text A}=0.05$ and near the order-disorder transition, both B and C
components self-assemble into two cylindrical domains which
distribute symmetrically in the space [see figure~\ref{fig3}~(a)].  The rest
of A component appears in the same region of B component as shown in
figure~\ref{fig3}~(c), since the
 volume fractions of A component  are too small to appear
alone in some space positions, where cylinders are similar to those
of  rod-coil diblock copolymers \cite{Chen2006}. When $\chi N$ 
increases, the A component is separated from the domain of B
component, and the lamellae become stable. In other words, the
cylinder/lamella transition appears with an increasing $\chi N$. The
reason for the formation of cylinders and the explanation for the
transition between cylinders and lamellae are in detail discussed by
Chremos and Theodorakis \cite{Chre2014}. They proposed that
a molecular asymmetry is required
for the formation of cylindrical domains, and the cylinder/lamella
transition results from the entropic penalty, which does apply to
diblock copolymers. When $f_{\text A}=0.05$, the coil-rod-coil triblock
copolymers are very asymmetric and are regarded as diblock copolymer, and
thus the cylinders form. The appearance of lamellae with $\chi N$ is
due to the behavior of the triblock copolymer. It is obvious that
herein the mechanism of the cylinder/lamella is different from the
one  proposed by Chremos and Theodorakis. The appearance of the
cylinder/lamella transition is to minimize the interaction energy
among different components, which is in reasonable agreement with
the same transition observed experimentally in rod-coil diblock
copolymers \cite{Shi2011}.

By contrast, when $f_{\text A}$ is large, only B component assembles into
cylinders, and the A and C components appear alternately in the
regions between the series of cylinders [see figure~\ref{fig3}~(b) and (c)].
When $\chi N$ is small, the interface between different component
domains is relatively thick, where the system tends to make more room
for the coil block to maximize entropy. When $\chi N$  increases,
the interface of lamellae tends to be thin to minimize the
interaction energies between different components, and to simultaneously give rise
to  the configuration entropic penalty. Compared
with lamellae, cylinders, where the rods are assembled into
interdigitated bilayer structure \cite{Chen2006}, are favorable to the
configuration entropy. Consequently, the cylinders are more stable
than lamellar phase for large $\chi N$, i.e., the lamella/cylinder
transition appears with $\chi N$.

\section{Conclusion and summary\label{sec4}}
Using the self-consistent field approach, the effect of asymmetry of
the coil block on the microphase separation is focused in ABC
coil-rod-coil triblock copolymers. The calculated results show that
the effect of the coil block fraction $f_{\text A}$ is dependent on $f_{\text B}$.
When $f_{\text B}$ is small, the effect of asymmetry of the coil block is
similar to that of  the  ABC flexible triblock copolymers; When
$f_{\text B}$ is an intermediate value, the self-assembly of ABC
coil-rod-coil triblock copolymers can behave like rod-coil diblock
copolymers under rather
asymmetrical conditions. When $f_{\text B}$ is large, the effect of asymmetry of the coil
block reduces. For  intermediate $f_{\text B}$, under the symmetrical and rather
asymmetrical conditions, an increase in the interaction parameter
between different components leads to  different transitions
between cylinders and lamellae.
In rod-coil ABC  systems, the asymmetric interaction
parameters can also be tuned up. It is expected that  the interesting morphologies can form
in the self-assembly under the conditions of asymmetric interactions, which is demonstrated by  the ABC coil-coil-coil
triblock copolymers \cite{Tang2004,Tyle2007}. This investigation will be
implemented in the subsequent work.

\section*{Acknowledgements} This research is financially supported
by the National Nature Science Foundations of China \linebreak (21564011), the
Nature Science Foundations of
the Inner Mongolia municipality (2017MS(LH)0211),\linebreak and
the  Graduate Education Innovation Program of the Inner Mongolia
Autonomous Province \linebreak(S20161012707).


\newpage
\ukrainianpart

\title{Вплив асиметрії виткоподібного блоку на самоскупчення у  ABC виток-стержень-виток триблокових кополімерах}
\author{К.-Г. Ґан\refaddr{label1}, Ґ.-Ґ. Менг\refaddr{label1}, Й.-Ґ. Ма\refaddr{label1}, С.-Л. Оуянг\refaddr{label2}}
\addresses{
\addr{label1} Вища школа природничих наук, Унiверситет науки i технологiй внутрiшньої Монголiї,\\ Баоту 014010, Китай
\addr{label2} Головна лабораторiя iнтегрованого використання мультиметалiчних ресурсiв Баян Обо, Унiверситет
науки i технологiї внутрiшньої Монголiї, Баоту 014010, Китай}

\makeukrtitle

\begin{abstract}
Використовуючи самоузгоджений польовий підхід, описано вплив асиметрії виткоподібного блоку на мікрофазне розділення в ABC
 триблокових кополімерах виток-стержень-виток. Для різних фракцій стержневого блоку  $f_{\text B}$, спостерігаються стійкі структури,
а саме ламели, циліндри, гіроїд і гексагональна ґратка сердечник-оболонка, а також побудовані фазові діаграми. Результати обчислень показують, що ефект фракції виткоподібного блоку $f_{\text A}$ залежить від $f_{\text B}$. 
Коли  $f_{\text B}=0.2$, тоді вплив асиметрії виткоподібного блоку є подібним до ефекту   ABC гнучких триблокових кополімерів. Коли $f_{\text B}=0.4$, 
тоді самоскупчення  ABC виток-стержень-виток триблокових кополімерів відбувається подібно до двоблокових кополімерів стержень-виток за певних умов. Коли $f_{\text B}$ продовжує зростати, тоді вплив асиметрії виткоподібного блоку зменшується. При  $f_{\text B}=0.4$, 
за симетричних та доволі асиметричних умов, збільшення параметра взаємодії між різними компонентами призводить до різних переходів між 
циліндрами та ламелами.
Результати вказують на деякий значний вплив ланцюжкової архітектури на самоскупчення, що може бути керівництвом для розробки та синтезу кополімерних матеріалів.

\keywords самоузгоджена польова теорія, стержень-виток блоковий кополімер, самоскупчення
\end{abstract}
\end{document}